\documentclass[a4paper,11pt]{article}
\pdfoutput=1 

\usepackage{jinstpub} 
\usepackage{graphicx}
\usepackage{subcaption}

\title{\boldmath The ATLASPix3.1 CMOS Pixel Sensor Testbeam Performance and Serial Powering Characterisation}

\author[a,1]{F. Ustuner,\note{Corresponding author.}}
\author[b,c]{R. Zanzottera,}
\author[b,c]{A. Andreazza,}
\author[e]{R. Dong,}
\author[d]{H. Fox,}
\author[a]{Y. Gao,}
\author[a]{P. Gheewalla,}
\author[a]{B. Masic,}
\author[d]{L. Meng,}
\author[e]{I. Peric,}
\author[b,c]{F. Sabatini}

\affiliation[a]{The University of Edinburgh,\\Edinburgh, The United Kingdom}
\affiliation[b]{INFN section of Milano\\Milano, Italy}
\affiliation[c]{Università degli Studi e INFN Milano\\Milano, Italy}
\affiliation[d]{Lancaster University,\\Lancaster, The United Kingdom}
\affiliation[e]{Karlsruhe Institute of Technology KIT,\\Karlsruhe, Germany}

\emailAdd{fuat.ustuner@cern.ch}

\abstract{High-voltage CMOS (HV-CMOS) pixel technology is being considered for future Higgs factory experiments. The ATLASPix3.1 chip, with a pitch of 50$\mu m$ x 150$\mu m$, fabricated using TSI 180nm HV-CMOS technology, is a full reticle-size monolithic HV-CMOS sensor with shunt-low dropout (LDO) regulators that allow serial powering for multiple sensors. A beam test was conducted at DESY using 3-6 GeV electron beams, with chips operated in triggerless readout mode with zero suppression, demonstrating multi-chip capability. This was further evaluated with hadron beams, both with and without the built-in power regulators. This study presents the electrical characterisations of the shunt-LDO regulators for serial powering and test beam results of ATLASPix3.1 sensors.}


\arxivnumber{2501.16887} 

\proceeding{Topical Workshop on Electronics for Particle Physics\\
  30 September- 4 October 2024\\
  Glasgow, The United Kingdom}

\begin{document}
\maketitle
\flushbottom

\section{Introduction}

High-voltage CMOS (HV-CMOS) sensors are a novel type of active pixel sensors for detecting ionizing particles, designed using CMOS processes with deep n-well options \cite{1}. The readout electronics are integrated within the pixel electrode, which collects electrons generated by traversing particles. This design isolates the electronics from the substrate, allowing the substrate to be biased with high voltage, enhancing sensor performance in terms of signal amplitude and charge collection speed \cite{1}. Additionally, their simpler production process and low production cost makes them suitable for high-energy physics experiments \cite{2}.

In this study, the performance of the ATLASPix3 HV-CMOS sensor has been evaluated in test beam experiments and local test benches for both multi-chip setups either powered in parallel or serial.

\section{ATLASPix3.1}

ATLASPix3.1 chip is the first full reticle-size monolithic HV-CMOS sensor with two shunt-LDO regulators. ATLASPix3.1 uses TSI 180 nm HV-CMOS process, and it has 200 $\Omega cm$ p-substrate with the deep n-well working as
the sensor electrode with a large fill factor \cite{3}. The size of the sensor is 2.0 x 2.1 $cm^2$ with a 150 x 50 $\mu m^2$ pixel pitch \cite{4}. The device features a breakdown voltage of approximately -60 V and supports timestamp (TS) with a precision of 25 ns. Additionally, it is capable of transmitting data off-chip at a maximum rate of 1.28 Gbit/s, enabling high-speed communication \cite{5}.

The matrix region comprises 132 columns, with each column containing 372 pixels. Each pixel is equipped with a custom design front-end, which includes a Charge Sensitive Amplifier (CSA), a comparator, a 3-bit threshold tuning Digital-to-Analog Converter (TDAC), a 4-bit Random Access Memory (RAM), and an output driver. Each column has a digital front-end that contains hit buffers (one per pixel), a 2x40 Content Addressable Buffers (CABs) and two End of Columns (EoCs) which provides both continuous and triggered readout. The system employs two distinct Readout Control Units (RCUs) for managing data acquisition modes: the RCU aux is dedicated to continuous readout, while the RCU main is responsible for handling triggered readout.

\section{The GECCO Readout System}

The Generic Configuration and Control (GECCO) system comprises hardware, firmware, and a suite of software tools. It was developed by the Karlsruhe Institute of Technology (KIT) to enhance the adaptability and flexibility of sensor test systems, providing a more flexible framework for various testing and configuration needs \cite{6}. 

\begin{figure}[htbp]
\centering
\includegraphics[width=0.7\textwidth]{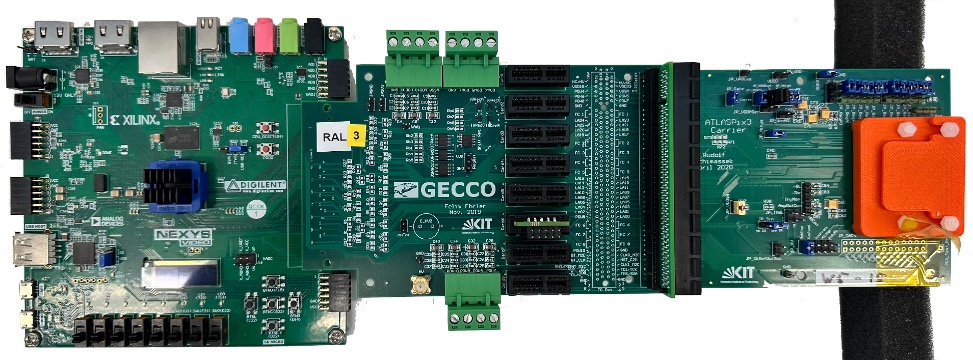}
\caption{\label{fig:1} The view of Generic Configuration and Control (GECCO) readout system.}
\end{figure}

Figure \ref{fig:1} illustrates the GECCO data acquisition (DAQ) system, featuring the Nexys Video Artix-7 FPGA and the ATLASPix3 Single Chip Carrier (SCC). The four-layer beam telescope and quad module (discussed in section \ref{sec:The Quad Module and Serial Powering (SP) Chain}) can be integrated into the DAQ system as replacement of the SCCs.

\section{Testbeam}

\begin{figure}[h]
\centering
\includegraphics[width=0.32\textwidth]{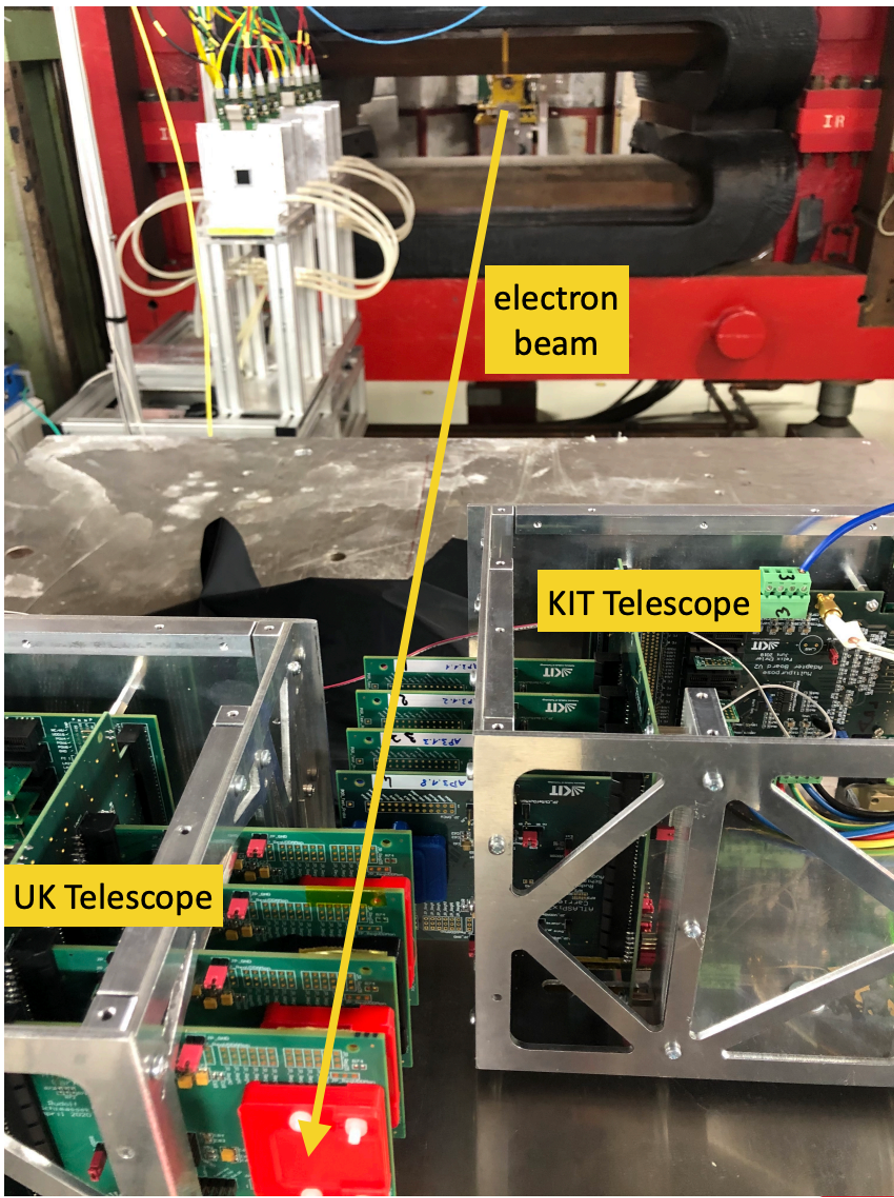}
\caption{\label{fig:2} The demonstration of ATLASPix3 four-layer beam telescopes from KIT (on the right) and UK (on the left) setup with beam direction. }
\end{figure}

The ATLASPix3 has been tested with electron beams energy range of 3-6 GeV at DESY \cite{7} in April 2022. The experimental setup is presented in Figure \ref{fig:2}.

The ATLASPix3 chips were operated in a hit-driven readout mode with zero suppression, and the beam was aligned perpendicular to the four-layer telescope setup. High-voltage (HV) scans for the telescope layers were performed in the range of -2V to -50V. The data reconstruction was carried out using the Corryvreckan framework which is modular and configurable software for reconstructing and analysing test beam and laboratory data \cite{9}. In this configuration, layer 3 served as the Device Under Test (DUT), while layer 1 was used as the reference. Layers 2 and 4 were employed for alignment purposes within the telescope system.

Crosstalk in silicon pixel detectors is unintended interference between neighboring pixels that can degrade signal fidelity and reduce resolution. In the case of the chips, the crosstalk between pixels, caused by the capacitive coupling of transmission lines between the pixel matrix and the EoCs readout, was effectively limited to approximately 1\% of the total hits.

\begin{figure} [htbp]
\centering
\subfloat[]{\includegraphics[width=0.4\textwidth]{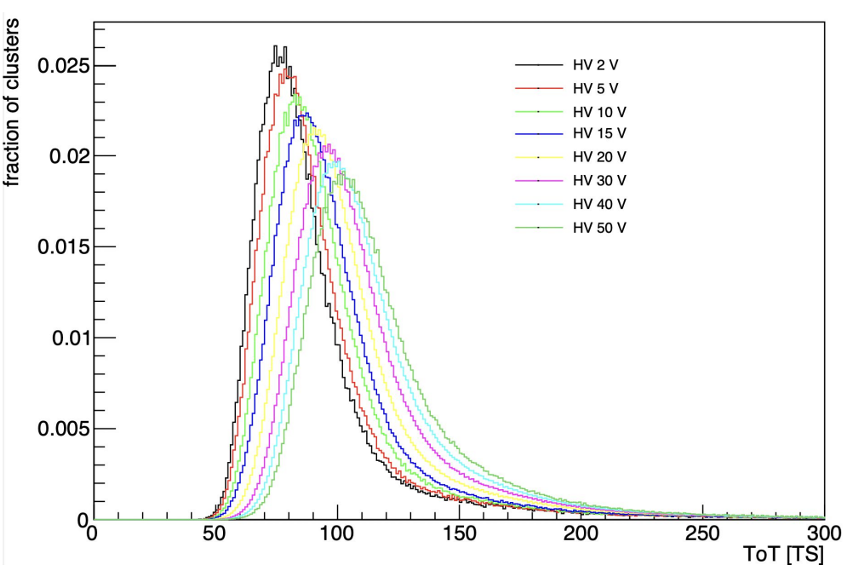} \label{fig:3a}}
\hspace{+0.5cm}
\subfloat[]{\includegraphics[width=0.4\textwidth]{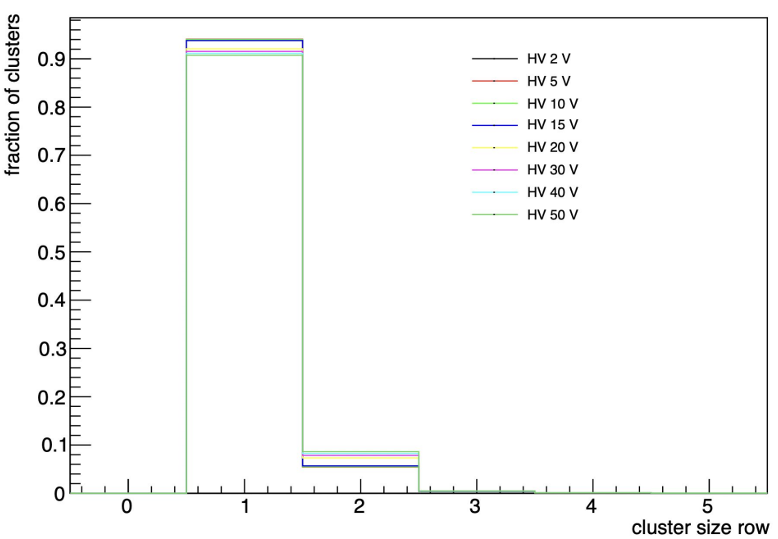} \label{fig:3b}}
\caption[Short caption]{(a) ToT distributions with high voltage range from -2V to -50V, where ToT is presented as unit in timestamp (TS) of 25 ns (b) the distribution of cluster size in row (the pitch of 50 $\mu m$) direction.}
\label{fig:3}
\end{figure}

Time-over-Threshold (ToT) distribution of the overlay of the all pixels is a form of Landau-Gaussian distribution as seen in Figure \ref{fig:3a}. The ToT distributions are proportional to increasing HV. For instance, the Most Probable Value (MPV) of ToT increases from 77 timestamps (TS) at -2V to 102 TS at -50V. A small amount of charge sharing is observed in the cluster size plot along the row direction as shown in Figure \ref{fig:3b}. The charge sharing is about 6\% at -2V while this value increases slightly by 2\% at -50V.

\begin{figure} [h]
\centering
\subfloat[]{\includegraphics[width=0.36\textwidth]{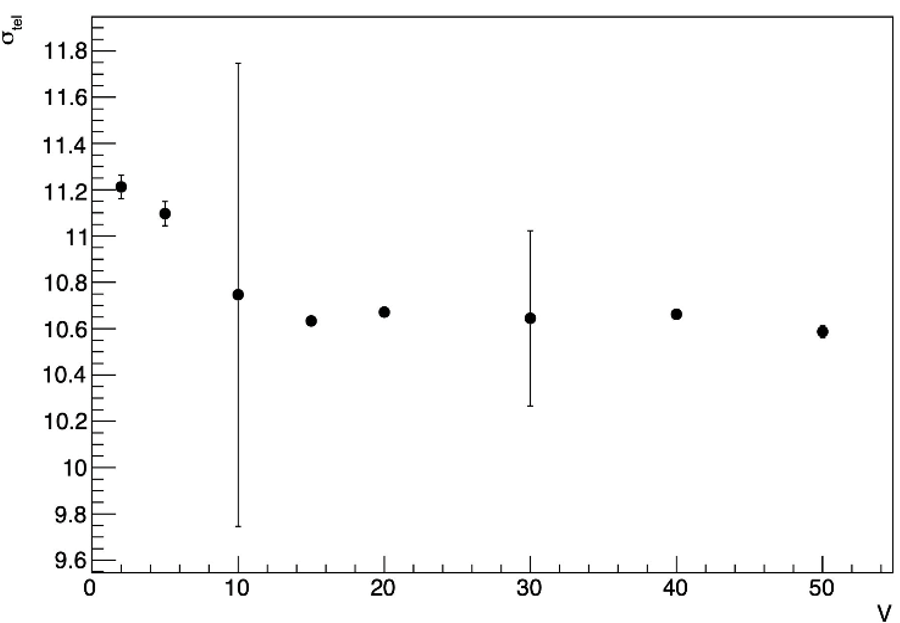}
\label{fig:4a}}
\hspace{+0.5cm}
\subfloat[]{\includegraphics[width=0.36\textwidth]{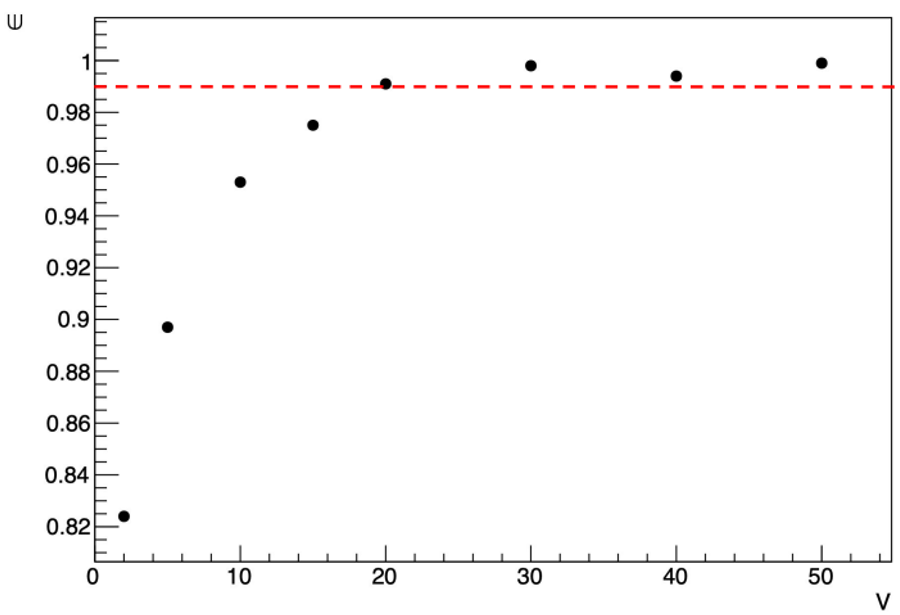}
\label{fig:4b}}
\caption[Short caption]{(a) The telescope spatial resolution and (b) efficiency values at different HV biases.}
\label{fig:4}
\end{figure}

Following the tests, the minimum spatial resolution achieved is 10.6 $\mu m$, as shown in Figure \ref{fig:4a}. This improvement is indicative of enhanced charge collection efficiency under higher bias conditions. Figure \ref{fig:4b} shows the distribution of the pixel efficiency maps at HVs. The efficiency reaches a plateau of around 99\% after -20V.

\section {The Quad Module and Serial Powering (SP) Chain}
\label{sec:The Quad Module and Serial Powering (SP) Chain}

\begin{figure} [htbp]
\centering
\subfloat[]{\includegraphics[width=0.2\textwidth]{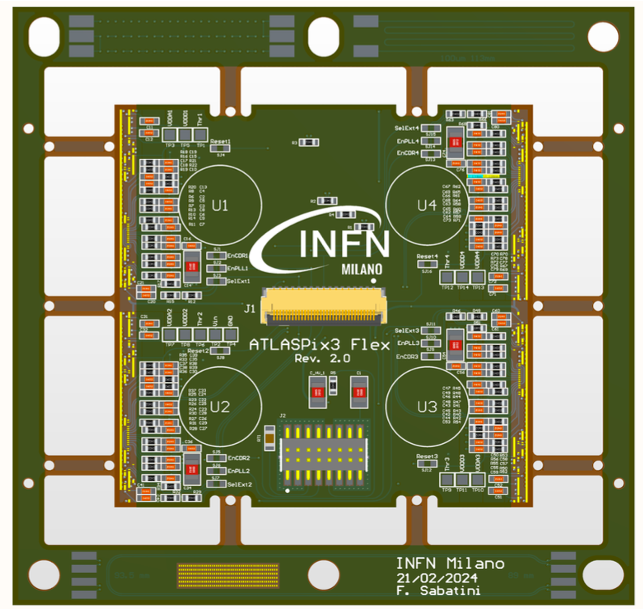}
\label{fig:5a}}
\hspace{+0.5cm}
\raisebox{0.1cm}{\subfloat[]{\includegraphics[width=0.6\textwidth]{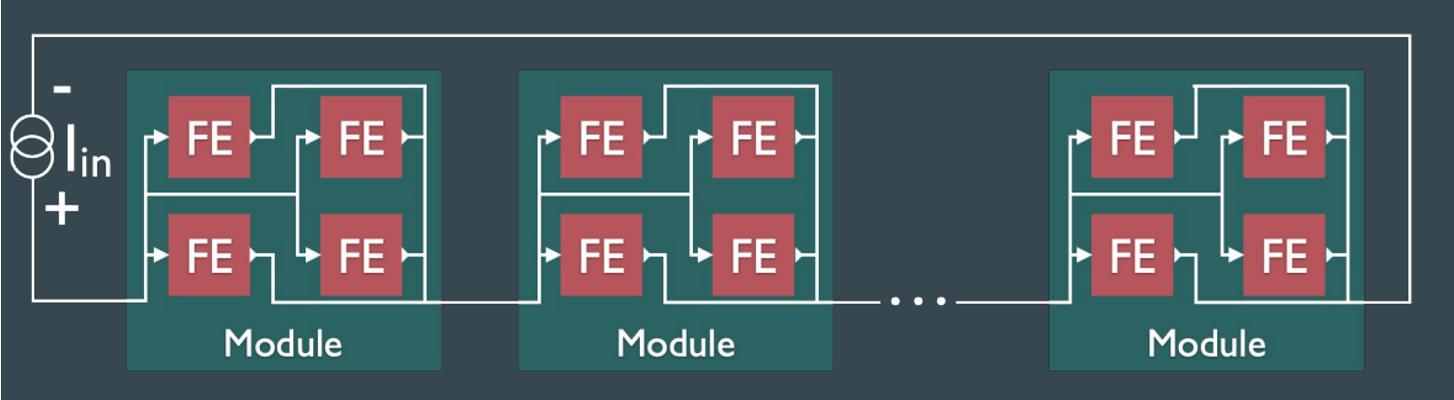} \label{fig:5b}}}
\caption[Short caption]{(a) The ATLASPix3.1 Quad Module (b) serial powering chain with quad modules \cite{8}.}
\label{fig:5}
\end{figure}

The multi-chip quad module system, comprising four ATLASPix3.1 chips within a 4x4 $cm^2$ area (see in Figure \ref{fig:5a}), facilitates shared powering (in parallel) and data transmission. In addition, serial powering involves the usage of a single, constant current source to operate the chips in a chain. This structure is aimed at minimizing power consumption, with a focus on efficiency and sustainability in electrical links. The multi-chip quad module approach and its integration of the serial powering scheme (see in Figure \ref{fig:5b}) are important for large-scale applications. 

\begin{figure} [htbp]
\centering
\subfloat[]{\includegraphics[width=0.4\textwidth]{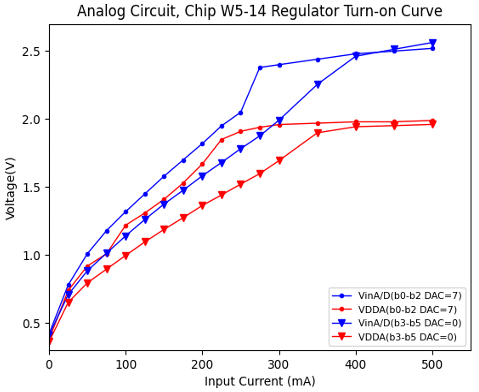}
\label{fig:6a}}
\hspace{+0.5cm}
\subfloat[]{\includegraphics[width=0.4\textwidth]{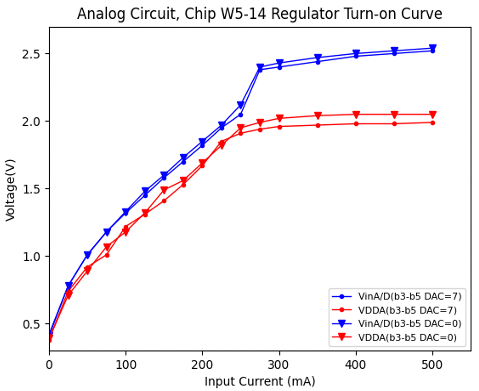}
\label{fig:6b}}
\caption[Short caption]{The shunt-LDO IV characteristics of VDDA and  Vin (a) by shunt 3-bit DAC values (b) by LDO 3-bit DAC values. VDDD presents a similar response.}
\label{fig:6}
\end{figure}

Two on-chip shunt-LDO regulators (for the supply voltage VDDD/A of the digital and analog parts of the chip) are implemented in ATLASPix3.1. Each regulator features a 6-bit DAC for two-step adjustments, 3 bits for the shunt regulator threshold and VDD tuning. Figure \ref{fig:6} presents the Current-Voltage (IV) characteristics of a single chip for both shunt and VDD tuning, using DAC values of 0 and 7. The current threshold is adjustable within  $\sim$ 100mA (Figure \ref{fig:6a}), while VDD has a limited tunability of about 0.15V (Figure \ref{fig:6b}). Additionally, ohmic behavior is observed beyond the threshold, with VDD regulated between 1.85V and 1.9V.

\begin{figure} [htbp]
\centering
\subfloat[]{\includegraphics[width=0.35\textwidth]{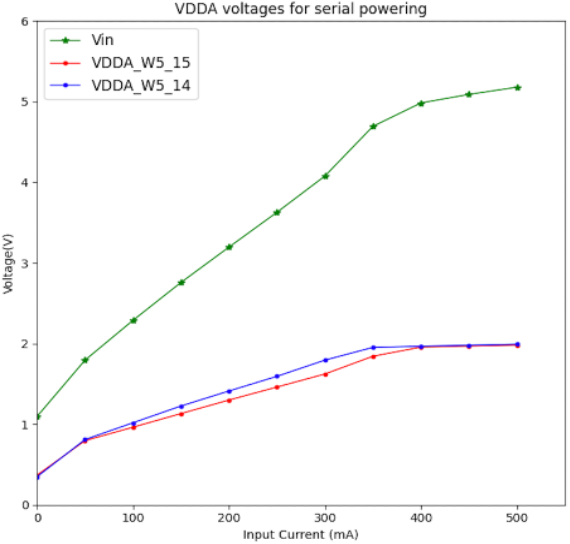}
\label{fig:7a}}
 \hspace{+0.5cm}
\subfloat[]{\includegraphics[width=0.35\textwidth]{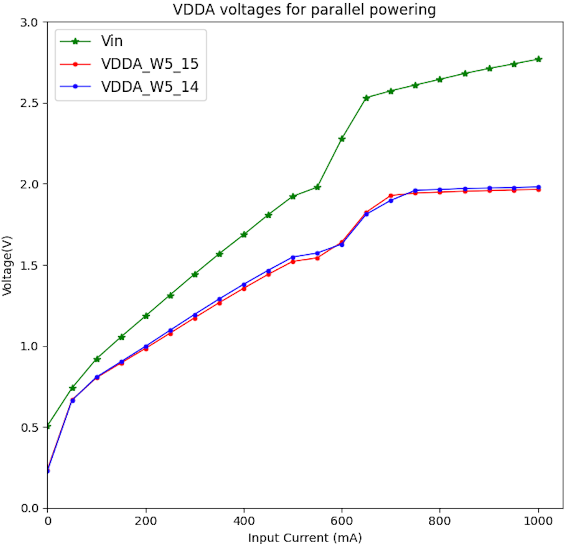}
\label{fig:7b}}
\caption[Short caption]{The shunt-LDO IV characteristics of VDDA with (a) the serial connection and (b) the parallel connection of two single chips. VDDD exhibits a similar response.}
\label{fig:7}
\end{figure}

Two single chips are then connected in both serial and parallel configurations to mimic the serial connection of quad modules and the parallel connection of chips within a quad module. In both setups, the VDD is regulated at 1.9V-1.95V as expected (see in Figure \ref{fig:7}). The same ohmic behavior is observed after the threshold. Subsequently, the chip operation and readout, including threshold and noise distribution measurements, were performed for both powering schemes. A minor variation of $\sim$8\% in the threshold and $\sim$5\% in noise distributions are observed between the different configurations. In these three cases the power consumption changes between $\sim$187$mW/cm^2$ to $\sim$237$mW/cm^2$.

\section{Conclusion and Further Plans}

The ATLASPix3 has been tested at DESY with 3-6 GeV electron beams with four-layer beam telescope. Overall pixel efficiency has been found to be around $99\%$, and the spatial resolution has been measured $10.6 \mu m$ applying -50V high voltage bias. 

The electrical characterisation tests of the shunt-LDO regulators on both single and multi-chip (parallel and serial connection) have been carried out. Output values of these regulators have been found to be at 1.85-1.95 V desired level. Moreover, the threshold and noise distribution variations across the three different setups are about 5\% and 8\%, respectively.  The short-term plan is to develop a serial powering demonstrator integrating several quad modules. These findings will be then used to provide inputs towards the design of large systems such as those planned in the HL-LHC upgrades in the LHCb experiment and future Higgs factory experiments.


\end{document}